\documentclass{elsart}
\def\be{\begin{equation}}
\def\ee{\end{equation}}
\def\bea{\begin{eqnarray}}
\def\eea{\end{eqnarray}}
 
\newcommand{\lb}{\label}
 
\usepackage{graphicx}

\usepackage{amssymb}

\begin{document}

\begin{frontmatter}


\title{Anomalous Response Variability in a Balanced Cortical Network Model}




\author{John Hertz}
\address{Nordita,
Blegdamsvej 17,
2100 Copenhagen, Denmark}
\ead{hertz@nordita.dk}

\author{Barry Richmond}

\address{Laboratory of Neuropsychology,
NIMH, Bethesda MD 20892 USA}
\ead{bjr@ln.nimh.nih.gov}

\author{Kristian Nilsen}

\address{Nordita,
Blegdamsvej 17,
2100 Copenhagen, Denmark}
\ead{kristnil@nordita.dk}

\begin{abstract}
 We use mean field theory to study the response properties of a simple
randomly-connected model cortical network of leaky integrate-and-fire neurons 
with balanced excitation and inhibition.  The formulation permits arbitrary 
temporal variation of the input to the network and takes exact account of 
temporal firing correlations.    We find that neuronal firing statistics 
depend sensitively on the firing threshold. In particular, spike count 
variances can be either significantly greater than or significantly less than 
those for a Poisson process.  These findings may help in understanding the 
variability observed experimentally in cortical neuronal responses. 
\end{abstract}

\end{frontmatter}

\section{Introduction}

Cortical neuronal responses often exhibit a puzzling variability (see, e.g.,
\cite{GWLR}):  spike count distributions obtained for repeated presentations
of a stimulus are frequently broader than Poisson distributions with the
same mean counts.  In this paper we study the statistics of neuronal responses  
in a simple model network with balanced excitation and inhibition and find
that they also show large variability.   We focus on the so-called Fano factor
$F$: the ratio of the variance of the spike count to its mean.  It is equal 
to 1 for any Poisson process, homogeneous or inhomgeneous.  We find that in 
our model it depends sensitively on the neuronal firing threshold: $F<1$ for high thresholds
and $F>1$ for low ones, while mean spike counts are hardly affected by changes
in threshold.    

To study the responses in our model, we utilize mean field theory, which is
exact in the limit of a large network with homogeneous connection probabilities 
\cite{Fulvi}.  It allows us to reduce the full network problem to one of a
single neuron subject to a Gaussian synaptic current, the mean and covariance 
of which are self-consistently related to those characterizing the the 
neuron's firing.  It is Gaussian because of the central limit theorem: it is 
the sum of a large number of (what can be proved to be) independent 
contributions from the other neurons.

The mean field theory of balanced cortical networks has 
been studied by a number of authors \cite{AmitBrunel,Brunel,vVS,Latham}, but 
generally in or near steady firing states and assuming that the self-consistent 
current input is uncorrelated.  Here, like ref.~\cite{Fulvi}, we consider 
time-dependent external drive (as in an experimental trial) and color the
noise correctly.

\section{The Model}

For these exploratory investigations, we use a network of $N$ mutually 
inhibitory neurons with a 10\% connection probability.  This 
is probably the simplest model for which one can achieve balanced asynchronous 
activity.  The nonzero connections are of equal strength, and the synaptic 
currents are assumed to act instantaneously, i.e., each presynaptic spike 
depresses the postsynaptic potential discontinously by a fixed amount.  We 
do not include transmission delays.  This and other features, such as the 
presence of two neuronal populations, inhibitory and excitatory, can easily 
be incorporated into the model.  A modest extension of this form might serve 
as a serious model of a cortical (mini)column.  However, we believe that the 
basic dynamical features of balanced asynchronous networks are already captured 
in our simple all-inhibitory model.  

We find it convenient to scale variables in the way used by van Vreeswijk and
Sompolinsky \cite{vVS}:  When the mean  number of neurons presynaptic to a
given one is $K \gg 1$, each synaptic strength is given a value 
$J/\sqrt{K}$.  Then the average synaptic current scales like 
$K \times 1/\sqrt{K} = \sqrt{K}$.  It is inhibitory and is counterbalanced  
by an external excitatory current which is also proportional to $\sqrt{K}$.  
The fluctuations in the synaptic current are smaller than the mean by a 
factor of order $1/\sqrt{K}$, i.e., of order 1.  In steady state, the two
currents nearly cancel, leaving a net current of order 1.  If the steady-firing
state is stable, disturbances of the rates relax very rapidly, in a time of
order $1/\sqrt{K}$.  While this picture is something of a caricature, we 
believe it has some relevance to cortical dynamics.

Mean field theory can be used for any kind of neuron, but here we use
leaky integrate-and-fire neurons.
Thus, the (subthreshold) membrane potentials are described by
\be
\dot u_i(t) = -\frac{1}{\tau}u_i + \sqrt{K}I^0_i(t) 
-\frac{J}{\sqrt{K}}\sum_{k, s} c_{ij} \delta(t-t^s_j),		\lb{eq:net}
\ee
where $\tau$ is their (common) membrane time constant, $I^0_i(t)$ is the
excitatory external input current felt by neuron $i$, $c_{ij}$ is 1 or 0
according to whether there is a connection from neuron $j$ to neuron $i$,
and $t^s_j$ is the $s$-th spike time of neuron $j$.  The external input
$I^0_i(t)$ represents input from elsewhere in the brain (e.g., the preceding
stage in a sensory pathway), and as such is noisy itself.  However, as the
recurrent connections in the randomly-diluted network generate dynamical
noise on their own, this extrinsic noise does not have a big qualitative
effect, so we take $I^0_i(t)$ to be constant here.  We also take it to be
uniform across the population.

\section{Mean Field Theory}

As indicated above, in mean field theory, one studies a single neuron for
which the recurrent synaptic current (the last term in (\ref{eq:net})) is
replaced by a self-consistent Gaussian current with self-consistent
mean and variance.  Explicitly, the subthreshold membrane potential of 
this single neuron obeys
\be
\dot u = -\frac{1}{\tau}u_i + \sqrt{K}[I^0 + I^1(t)]. 		\lb{eq:mf}
\ee
Whenever $u$ reaches a threshold $\theta$, it fires a spike and $u$ is 
reset to zero.  The effective recurrent current $I^1(t)$ has a mean
\be
\langle I^1(t) \rangle = -J r(t) 				\lb{eq:mn}
\ee
proportional to the instantaneous firing rate $r(t)$ of the neuron and a
covariance
\be
\langle \delta I^1(t) \delta I^1(t') \rangle 
= \frac{1-K/N}{K} C(t,t'),				\lb{eq:var}
\ee
where $C(t,t') = \langle [S(t) - r(t)][S(t')-r(t')]\rangle$ 
is the autocorrelation funtion of the neuronal firing $S(t) 
=\sum_s \delta(t-t^s)$.  The $K$ in the denominator  in (\ref{eq:var}) 
comes from the averaging over $K$ independent inputs, and the $1-K/N$
in the numerator is a correction for finite connection concentration $K/N$.  
It can be derived using the methods of ref.~\cite{KreeZip}.

This model can not be solved exactly analytically, but it is simple to
solve numerically, using the method first introduced for spin glasses by
Eisfeller and Opper \cite{EisOpp}.  We start with a guess at the form of
the mean and covariance function of the random current $I^1(t)$ and run a 
series of ``trials'', in each of which we integrate (\ref{eq:mf}) for an 
independent realization of $I^1(t)$.  By averaging the output of the neuron 
over the trials, we get an estimate of $r(t)$ and $C(t,t')$, which is used
to generate new examples of $I^1(t)$ for another set of trials.  This loop
is then iterated until the statistics converge to self-consistency.  In our 
calculations here, we used 10000 trials per iteration and up to 30 iterations.

Each trial was 100 integration steps (which we call ``milliseconds'')
long.  We chose parameters $K=500$, $N=5000$, and $\tau = 10$ ms.  The external 
excitatory input $I^0(t)$ was constant at a low value (evoking a background 
firing rate of 3 Hz) during the first and last 10 ms. In the middle 
80 ms, an additional ``stimulus'' input, which peaked about 15 ms after onset, 
was added.  It typically evoked about 4-5 spikes, with peak rates around 100 
Hz.  The spike count distributions, PSTHs and covariance functions were
computed for a number of values of the firing threshold $\theta$.

\begin{figure}[htbp]
\includegraphics[width=10cm,height=14cm,angle=270]{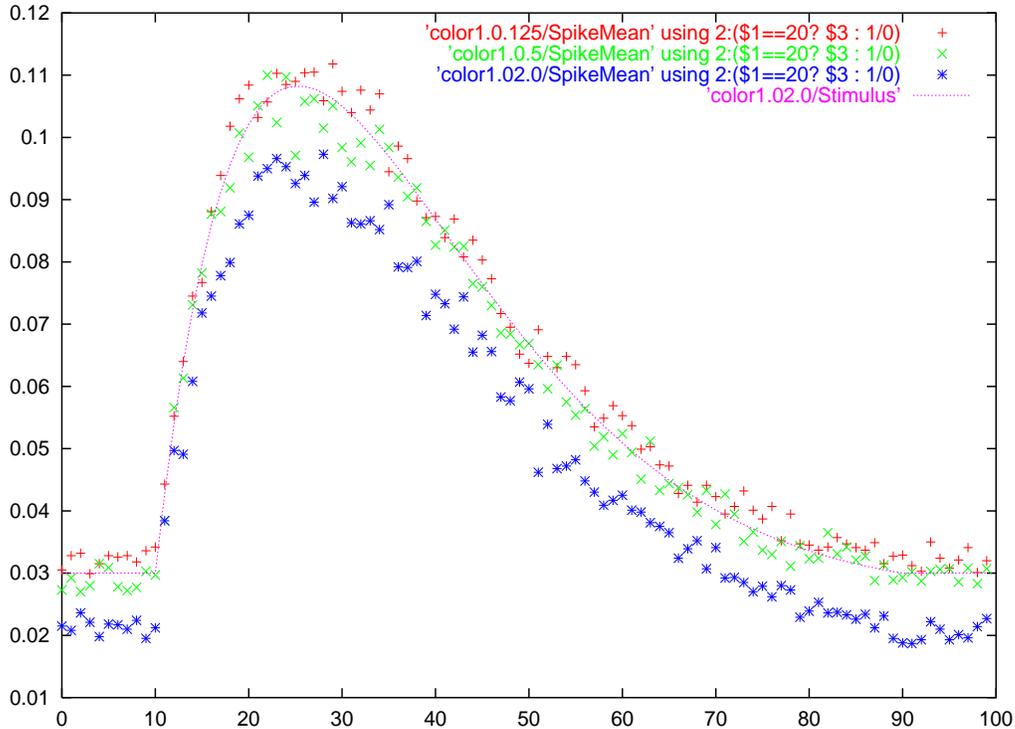}
\caption{Stimulus (solid line) and PSTHs for three threshold values:
0.125 (red), 0.5 (green) and 2.0 (blue).}
\label{fig:psths}
\end{figure}

\begin{figure}[t]
 \includegraphics[width=10cm,height=14cm,angle=270]{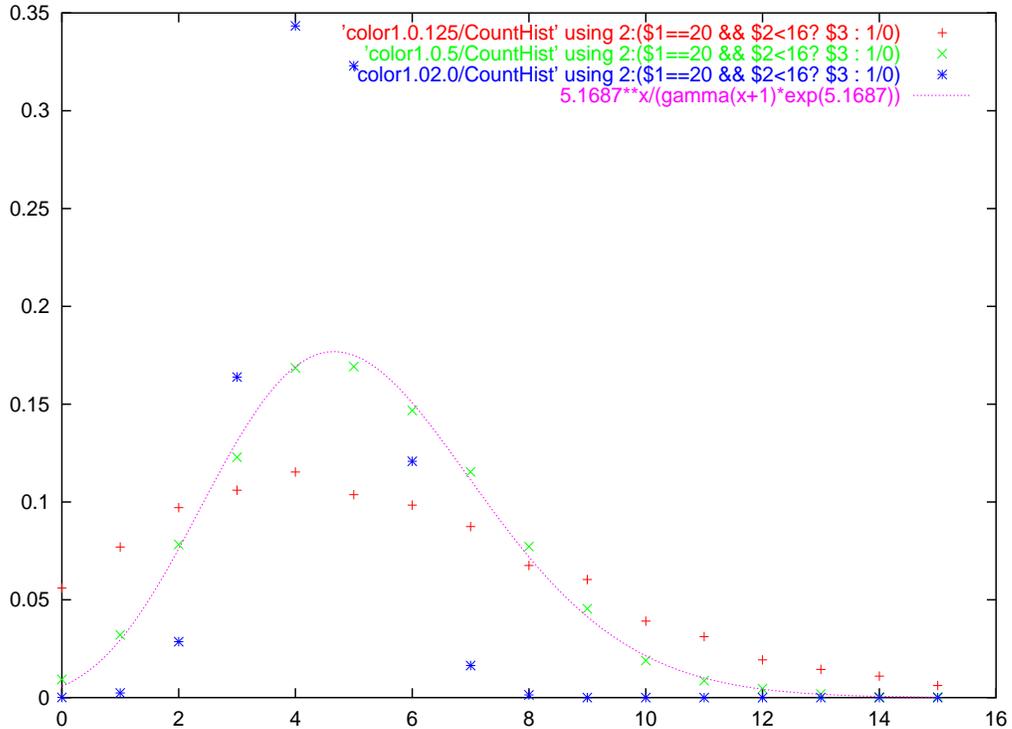}
\caption{Count statistics for threshold  0.125 (red), 0.5 (green)
and 2.0 (blue).  The solid line shows the good fit to a Poisson 
distribution for the mean spike count (5.1687) obtained for threshold 0.5.} 
\label{fig:counts}
\end{figure}

\begin{figure}[t]
\includegraphics[width=10cm,height=14cm,angle=270]{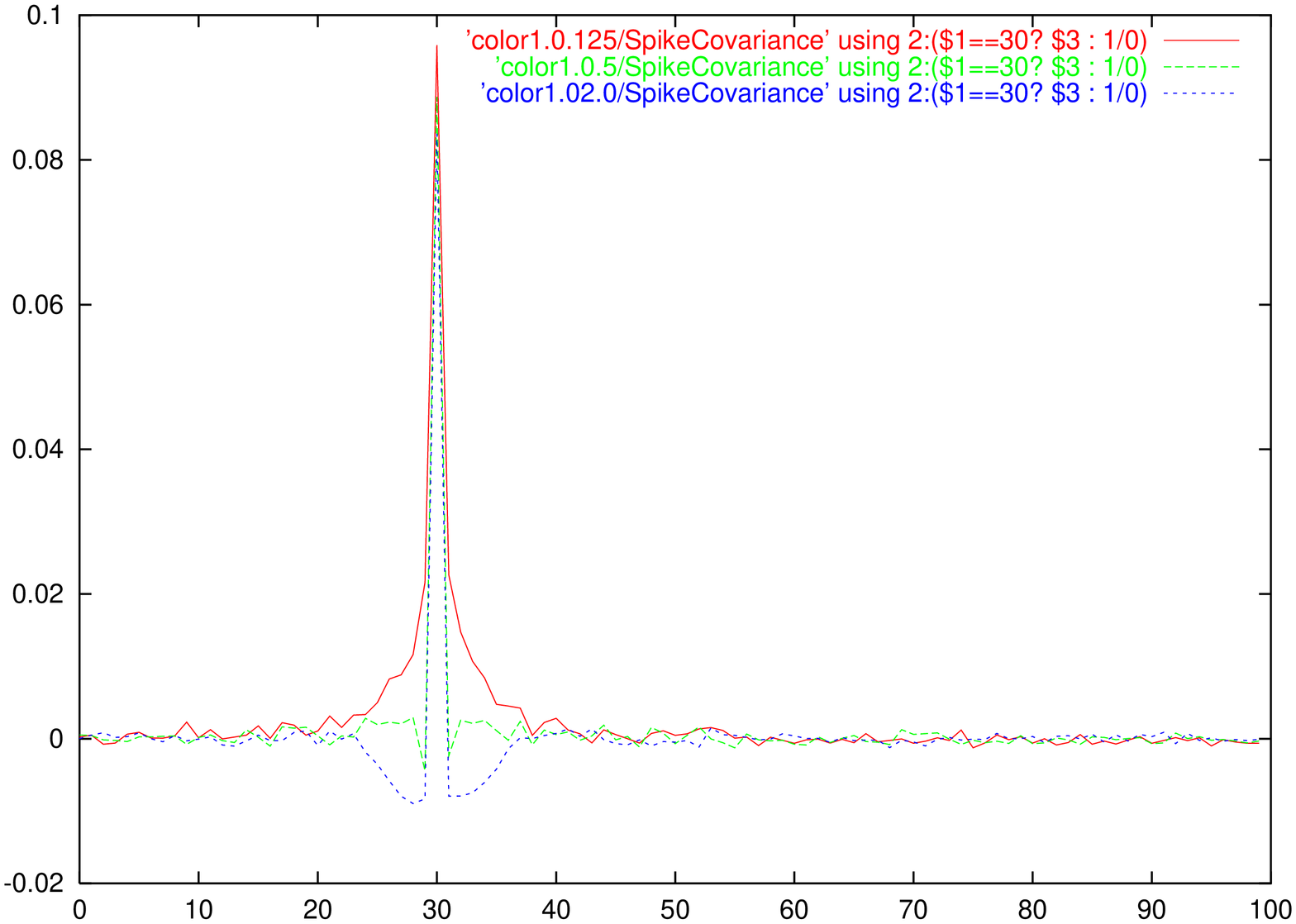} 
\caption{Autocorrelation functions $C(30,t')$ for thresholds
0.125 (red), 0.5 (green)
and 2.0 (blue).} 
\label{fig:covar}
\end{figure}

\section{Results}

For all cases studied, the relaxation of the network to its state of 
balanced excitation and inhibition was very rapid, as expected, so the
response tracked the time course of the excitatory external input 
closely (Fig.~\ref{fig:psths}).  The overall response strengths vary
only weakly with threshold: a factor of 16 difference between the smallest
and largest threshold values produced only a 20\% difference in mean response.
This is because the increased firing that would be produced by lowering the
threshold is largely compensated by the comcomitant increase in inhibition.

However, varying the threshold had a strong effect on the irregularity of
the firing.  Fig.~\ref{fig:counts} shows the spike count distributions for
three threshold values with ratio 1:4:16.  While the intermediate value fits
a Poisson distribution well, the low threshold leads to an anomalously broad
distribution ($F=2.4$) and the high one to an anomalously narrow one
($F = 0.25$).

These differences are also evident in the autocorrelation function $C(t,t')$.
Fig.~\ref{fig:covar}) shows $C(t,t')$ as a function of $t'$ for a fixed 
value of $t$ for the same three threshold values as in the preceding figures.
For the lowest threshold, there is a ``hill'' centered around $t=t'$, while
for the highest there is a valley.  These are indicative of spike ``bunching''
and ``antibunching'', respectively, leading naturally to the higher and lower
spike count fluctuations seen in Fig.~\ref{fig:counts}. The intermediate 
threshold value shows very little correlation (apart from the delta-function 
peak at $t=t'$), consistent with the nearly-Poisson count distribution found 
in this case.

\section{Discussion}

Measured Fano factors in visual and IT cortex \cite{GWLR} vary over a range 
at least as large as the one-order-of-magnitude difference between those
for the smallest and largest thresholds described above.  Of course, threshold
differences are not the only possible source of such response variability.  
We have also explored the effects of varying synaptic strengths, with similar
results, and it seems likely that diffferences in a wide variety of 
single-neuron properties can have the same kind of effect. 

Neurons in a local cortical network can not all be expected to have the same 
threshold, and, furthermore, their thresholds (or other parameters) may 
fluctuate (uncontrollably) from trial to trial.  We have also found large 
variations in the Fano factor in a model where these fluctuations are assumed
independent for different neurons and in different trials.  

All these results are only suggestive, and more systematic work, both 
experiments and modeling, is called for.  However, they do point to the
possibility that the observed response variability of cortical neurons
may be accounted for in terms of natural variations in properties from
neuron to neuron and trial to trial.

\end{document}